\documentclass{article}

\usepackage{graphicx}
\usepackage{bm}
\usepackage{amsmath}
\usepackage{amssymb}
\usepackage{url}

\usepackage[margin=1in]{geometry}

\usepackage[hidelinks]{hyperref}
\usepackage{breakurl}

\begin{document}

\title{\bf Taking a Lesson from Quantum Particles \\ for Statistical Data Privacy}

\date{}

\author{Farhad Farokhi\thanks{F. Farokhi is with the CSIRO's Data61, Docklands, Australia and the University of Melbourne, Parkville, Australia. e-mail: farhad.farokhi@unimelb.edu.au}}

\maketitle

\begin{abstract}
Privacy is under threat from artificial intelligence revolution fueled by unprecedented abundance of data. Differential privacy, an established candidate for privacy protection, is susceptible to adversarial attacks, acts conservatively, and leads to miss-implementations because of lacking systematic methods for setting its parameters (known as the privacy budget). An alternative is information-theoretic privacy using entropy with the drawback of requiring prior distribution of the private data. Here, by using the Fisher information, information-theoretic privacy framework is extended to avoid unnecessary assumptions on the private data. The optimal privacy-preserving additive noise, extracted by minimizing the Fisher information, must follow the time-independent Schr\"{o}dinger's equation. A fundamental trade-off between privacy and utility is also proved, reminiscent of the Heisenberg uncertainty principle.
\end{abstract}

\section{Introduction}
Privacy, defined as control over knowledge about oneself~\cite{fried77charles}, is an important part of personal liberty, moral autonomy, and democracy~\cite{mcneil2011privacy,gavison1980privacy}. This notion has been increasingly under threat by artificial intelligence in the era of big data~\cite{ICT4DBibliography2410,mcneil2011privacy}, which motivates the need for development of methods for preserving privacy of individuals. The roll out of the General Data Protection Regulation (GDPR), the legal text of the Regulation (EU) 2016/679, will undoubtedly further fuel the privacy research by industry so that corporations can recruit and maintain a larger set of customers willing to share their private data under proper guarantees.

Researchers at Microsoft developed the notion of differential privacy in mid-2000s as a methodology for responding to online queries on private datasets in a privacy-preserving manner~\cite{dwork2010differential,10100711681878_14}, following negative results establishing impossibility of publishing outcomes of arbitrary aggregative queries on private databases without revealing private sensitive information~\cite{dinur2003revealing,denning1979tracker}.
Differential privacy, roughly speaking, requires that the output a query on a dataset should be systematically corrupted, for instance by an additive noise, so that its statistics do not significantly change by changing the private data of one individual; hence rending the problem of reverse engineering the private dataset over which the queries were computed hopeless.

 Differential privacy has proved to be an extremely popular methodology~\cite{dwork2014algorithmic}, even extending beyond research to government organizations and private corporations~\cite{appledifferentialprivacy,abowd2018us}. Even ignoring susceptibility of differential privacy to adversarial attacks~\cite{haeberlen2011differential}, the notions is fairly conservative~\cite{bambauer2013fool,1010079783642158384_18,garfinkel2018issues}. Further, because of the lack of a systematic way for setting the so-called ``privacy parameter'', differential privacy has shown to be ineffective in practice~\cite{appleprivacyproblems,tang2017privacy}.

Another approach for ensuring privacy is to use information or estimation theory metrics in order to measure private information leakage when responding to queries on private datasets~\cite{liang2009information,braca2016learning,liu2017information}. These studies date back to wiretap channels~\cite{6772207} and their extensions~\cite{yamamoto1983source}. Information-theoretic measures of privacy rely on mutual information (or relative entropy) for measuring private information leakage and cast the privacy problem as a generalized rate-distortion problem~\cite{liang2009information}. Although possessing physical interpretations for privacy and offering strong guarantees, the use of entropy as a measure of privacy forces the private dataset to be statistically distributed with known distributions. This is the biggest shortfall of information-theoretic privacy because it introduces information asymmetry between the adversary and the curator (denoting those who are burdened with protecting data privacy) with potential for devastating mistakes by the curator due to misinformation. 

This motivates the extension of information-theoretic privacy to use the Fisher information, from the estimation theory, as measure of private information leakage. This is because the Fisher information does not require strong statistical assumptions on the private dataset.

\section{Problem Formulation}
Here we model a dataset with vector $x\in\mathbb{R}^n$. Assume that a $m$-dimensional query $f:\mathbb{R}^n\rightarrow\mathbb{R}^m$ is provided to the curator for execution on the dataset $x$. The curator, instead of directly releasing the outcome $f(x)$, provides a noisy response as in
\begin{align} \label{eqn:measurement}
y=f(x)+w,
\end{align}
where $w$ is an additive noise with density function $p_w:\mathcal{W}\rightarrow \mathbb{R}$ with $\mathcal{W}\subseteq\mathbb{R}^m$ denoting the support set of the additive noise. The support set can capture the noises that are sensible in practice in order to avoid absurd outputs. This is a feature that is entirely missing from differential privacy~\cite{bambauer2013fool}. The Cram\'{e}r-Rao bound~\cite{bercher2012generalized, cramerraotheorem, hero1996exploring} shows that the quality of an adversary's estimate of the private dataset $x$ based on measurement $y$ in~\eqref{eqn:measurement} denoted by $\hat{x}(y)$ is bounded by
\begin{align} \label{eqn:bound}
\mathbb{E}\{\|x-\hat{x}(y)\|_2^2\}\geq \frac{\alpha(x)}{\mathfrak{I}},
\end{align}
where $\alpha(x)$ is a term that is a function of the Jacobian of $f(x)$ and $\mathfrak{I}$ is the Fisher information of the additive noise defined as
\begin{align}
\mathfrak{I}=\mathbb{E}\left\{\left\|\frac{\partial}{\partial w}\log(p_w(w))\right\|_2^2 \right\}.
\end{align}
This bound is valid of unbiased estimators, $\mathbb{E}\{\hat{x}(y)\}=x$, but it can be easily generalized to biased estimators with some additional terms~\cite{bercher2012generalized, cramerraotheorem}. Intuitively, the most private action can be extracted by maximizing the adversary's estimation error $\mathbb{E}\{\|x-\hat{x}(y)\|_2^2\}$ but this is a function of the actions of the adversary, namely its estimator $\hat{x}(y)$. In a real world, we do not know the adversary and thus need to be prepared for the worst-case scenario. Therefore, minimizing $\mathfrak{I}$ can capture the most privacy-preserving additive noise. This is a great measure of privacy because it is independent of the actions of the adversary and only depends on the choice of the additive privacy-preserving noise. 

This framework can also be made mindful of the utility of the response by enforcing a constraint on 
\begin{align}
\mathfrak{Q}=\mathbb{E}\left\{g(w) \right\},
\end{align}
where $g:\mathbb{R}^m\rightarrow \mathbb{R}_{\geq 0}$ is increasing in each component of $w$; therefore, the quality of response is inversely proportional to $\mathfrak{Q}$. For instance, $g(w)=w^\top w$ selects the trace of the co-variance matrix of the additive noise to be inversely proportional to the quality or utility of response but the choice is not limited to  co-variance. 

Now, we can pose the problem of optimal privacy-preserving policies as
\begin{subequations}
\begin{align}
\min_{\gamma}\quad\;   & \quad  \mathfrak{I}, \\
\mathrm{such\;that}  &\quad  \mathfrak{Q}\leq \rho.
\end{align}
\end{subequations}
The constraint $\mathfrak{Q}\leq \rho$ ensures that the quality of the privacy-preserving noise can be set in advance, which is not possible in differential privacy.

\section{Privacy and the Schr\"{o}dinger's Equation}
To solve this optimization problem using the method in~\cite{jeyakumar1990zero}, we can form the Lagrangian
\begin{align*}
\mathcal{L}=&\int_{\mathcal{W}} \left\|\frac{\partial}{\partial w}\log(p_w(w))\right\|_2^2p_w(w) \mathrm{d}w+\lambda\left(\rho-  \int_{\mathcal{W}}g(w)p_w(w)\mathrm{d}w\right)+\mu\bigg(\int_{\mathcal{W}}p_w(w)\mathrm{d}w-1\bigg),
\end{align*}
where $\lambda\leq 0$ is the Lagrange multiplier corresponding to the inequality constraint $\mathfrak{Q}\leq \rho$ and $\mu\in\mathbb{R}$ is the Lagrange multiplier corresponding to the fact that the integration of the density $p_w(w)$ over $\mathcal{W}$ is equal to one (because it is a probability density function). Using Theorem~5.3 in~\cite{edwards1973advanced}, the necessary condition for minimizing the Lagrangian $\mathcal{L}$ is that
\begin{align} \label{eqn:schrodinger}
\nabla^2 \psi(w)-\frac{1}{4}\bigg(\mu+\lambda g(w) \bigg)\psi(w)=0,\quad \forall w\in\mathcal{W},
\end{align}
where $\psi(w)=\sqrt{p_w(w)}$. 

The optimality condition in~\eqref{eqn:schrodinger} is the non-relativistic time-independent Schr\"{o}dinger's equation with potential function proportional to $\mu+\lambda g(w)$ inside the set $\mathcal{W}$ and infinite potential outside. Note that~\eqref{eqn:schrodinger} can essentially be rewritten in the more familiar form of
\begin{align} 
\boxed{\left[-\nabla^2 +V(w)\right]\psi(w)=0,}
\end{align}
where 
\begin{align}
V(w)=
\begin{cases}
(\mu+\lambda g(w))/4, & w\in\mathcal{W},\\
+\infty, & w\notin \mathcal{W}.
\end{cases}
\end{align}

Minimization of the Fisher information and its relationship with the Schr\"{o}dinger equation was previously explored in~\cite{frieden1990fisher,reginatto1998derivation}. However, in those papers, the choice of Fisher information as a measure of disorder in quantum mechanics was merely philosophical while, here, the choice is concretely motivated by the problem of privacy.

As an example, consider $m=1$ with $\mathcal{W}=[-a,a]$ and assume that $g(w)=0$. This is identical to the problem of a free quantum particle in an infinite square well. The solutions to the time-independent Schr\'{o}dinger equation in this case is known to be of the form
$\psi(w)=c\sin(n\pi(w-a )/(2a))$ for all $n\in\mathbb{N}$. Note that $c=1/\sqrt{a}$ because the integration of the density $p_w(w)=\psi(w)^2$ over $\mathcal{W}$ is equal to one. Hence, 
$p_w(w)=\sin^2(n\pi(w-a )/(2a))/a$ for all $n\in\mathbb{N}$. As a result, it can be seen that
$\mathfrak{I}=n^2\pi^2/a$ for all $n\in\mathbb{N}$. As it is desired to minimize $\mathfrak{I}$ for ensuring the most privacy, it must be that $n=1$ for the optimal privacy-preserving policy. Another example, is to consider unbounded $\mathcal{W}=\mathbb{R}$ and assume that $g(w)=w^2$. In this case, simple algebraic derivations show that the optimal privacy-preserving policy is $
p_w(w)=\exp(-w^2/(2\rho))/\sqrt{2\pi\rho},$ which is the Gaussian density function that is known to guarantee a relaxed version of differential privacy, known as the approximate differential privacy~\cite{10100711761679_29,dwork2014algorithmic}.

Finally, we would like to prove a fundamental result regarding the privacy-utility trade-off reminiscent of the Heisenberg uncertainty principle. This is only shown for $g(w)=w^\top w$. It is known that $\mathfrak{J}\geq 1/\mathbb{E}\{w^\top w\}$ for any density function $p_w$~\cite{stein2014lower} and, as a result,
\begin{align} \label{eqn:tradeoff}
\boxed{\mbox{Privacy Principle:}\quad \mathfrak{J}\mathfrak{Q}\geq 1.}
\end{align}
This inequality proves that, by decreasing $\mathfrak{J}$ in order to improve privacy, the quality must be sacrificed by increasing $\mathfrak{Q}$. This is underlying principle in privacy: there is no free lunch for privacy.

Finally, note that, although previously, the Fisher information has been used for measuring privacy infringements in~\cite{farokhi2018fisher}, parallels with quantum physics were not explored and the fundamental result regarding the privacy-utility trade-off in~\eqref{eqn:tradeoff} was also not investigated.

\section{Conclusions}
In this paper, we used the Fisher information as a measure of privacy to avoid unnecessary assumptions on the private data. We proved that the optimal privacy-preserving additive noise, extracted by minimizing the Fisher information, must follow the time-independent Schr\"{o}dinger's equation. We also derived a fundamental trade-off between privacy and utility.

\bibliographystyle{plain}
\bibliography{sample}

\end{document}